\documentclass[fleqn,usenatbib]{mnras}

\usepackage{newtxtext,newtxmath}

\usepackage[T1]{fontenc}

\DeclareRobustCommand{\VAN}[3]{#2}
\let\VANthebibliography\thebibliography
\def\thebibliography{\DeclareRobustCommand{\VAN}[3]{##3}\VANthebibliography}

\usepackage{graphicx}	
\usepackage{amsmath}	

\title[NGC 6139 their reddening and membership]{The variable stars in the field of NGC 6139: \\ A critical approach to their reddening and membership.}

\author[Yepez et al.]{
M. A. Yepez$^{1}$\thanks{E-mail: myepez@astro.unam.mx}, A. Arellano Ferro$^{1}$, I. Bustos Fierro$^{2}$, A. Luna$^{3,4}$
\\
$^{1}$Instituto de Astronom\'ia, Universidad Nacional Aut\'onoma de M\'exico, Ciudad de M\'exico, CP 04510, M\'exico.\\
$^{2}${Observatorio Astronómico, Universidad Nacional de Córdoba, Córdoba C.P. 5000, Argentina.}\\
$^{3}${Instituto de Astrofísica, Facultad de Ciencias Exactas, Universidad Andrés Bello, Fernández Concha 700, Las Condes, Santiago, Chile.}\\
$^{4}${European Southern Observatory, Karl-Schwarzschild-Straße 2, 85748, Garching, Germany}\\
}

\date{Accepted -- 27. Received --; in original form --}

\pubyear{2020}

\begin{document}
\label{firstpage}
\pagerange{\pageref{firstpage}--\pageref{lastpage}}
\maketitle


\begin{abstract}
We present a CCD $VI$ time-series analysis of the globular cluster NGC 6139 and its variable star population. Using the astrometric data available in $Gaia$-DR3 we performed a membership analysis that enabled the construction of a clean Colour-Magnitude Diagram (CMD). Variable stars in the field of the cluster reported by $Gaia$-DR3 and newly discovered ones in this paper are classified and their membership is critically evaluated. We report two cluster member RRc (V12, V15) and four SR (V13, V14, V17, V18) not previously detected and assign variable names to  V11 and V16 detected by $Gaia$ as they proof to be cluster members. Light curves and periods for non-member $Gaia$ eclipsing binaries, semi regular variables and newly detected RR Lyrae stars are provided. Fourier decomposition of the light curves of the cluster member RRab and RRc stars leads to the values [Fe/H]=$-1.63$ dex, and distance of 9.63$\pm$0.68 kpc. The Oosterhoff type II nature of the cluster is confirmed. We adopted the mean reddening $E(B-V)$=0.786 mag and performed a differential reddening analysis based on the dispersion of the red giant branch. The differential map allowed a mild correction of the CMD.
\end{abstract}



\section{Introduction}

NGC 6139 is a Galactic globular cluster in the constellation of Scorpius, located at $\alpha = 16^{\rm h} 27^{\rm m} 40.37^{\rm s}$, $\delta = -38^{\circ} 50' 55.5''$ (J2000) and very near to the Galactic bulge at $l$=342.37$^{\circ}$, b=6.94$^{\circ}$, hence it is subject to a large reddening very likely of differential nature. In the compilation of  \citet{Harris1996} (Edition 2010), NGC 6139 is listed with a distance $d=10.1$ kpc, a metallicity $[\rm{Fe/H}]=-1.65$ dex and a mean reddening $E(B-V)=0.75$ mag. The dust calibration of \citet{Schlafly2011}, for the cluster position renders values of $E(B-V)$ between 0.76 and 0.82 mag and a mean of $0.786 \pm 0.017$ mag.

NGC 6139 is a poorly studied cluster. For more than 30 years an analysis of its variable stars has not been carried out. \citet{Hazen1991} using 32 plates in the B passband, found ten and six variable stars inside and outside of the tidal radius of the cluster, respectively. She suggested that the ten inner variables are probably members of the cluster. These ten variables are the only ones listed in the Catalogue of Variable Stars in Globular Clusters (CVSGC) in its 2013 edition \citep{cle01}. \citet{Hazen1991} does not give coordinates for the 10 variables, but she reported periods for the five variables labeled as RR Lyrae stars. In the $\log$ P vs $A_B$ plane of her figure 5 it is clear that the five RR Lyrae stars follow the distribution typical of a Oosterhoff type II
(Oo II) cluster. The author found the mean $V$-band mean magnitude of the horizontal branch (HB) to be $V_{HB}= 17.8 \pm 0.4$ mag for a value of $E(B-V)=0.74$ mag.

\citet{Samus1996} gave a color magnitude diagram with $B$ and $V$ photometry and provided the equatorial coordinates for the ten variables star in NGC 6139. These authors estimated $V_{HB}=18.0$ mag, and by assuming an age of 14 Gyrs and reddening between 0.82 and 0.87 mag, concluded that [Fe/H] $\leq$ 2.0 dex. In the last two decades of the last century, several indepenedent estimates of the reddening, distance and metallicity of NGC 6139 were performed by numerous authors. These values range 0.68-0.87 mag, 9.4-10.1 kpc and -1.71 to -1.28 dex respectively . A detail comparison of all these determinations and our own,  will be tabulated towards the end of the present work. Finally, \citet{Samus2009} refined the coordinates of the ten variables stars, which were adopted by Clement in the CVSGC.

While the above mentioned determinations of the reddening, distance and metallicity are rather in agreement, within the uncertainties, a time-series analysis of the variable star population of NGC 6139 seems to be in order in the CCD plus differential imaging approach era. We are interested in framing NGC 6139 in the homogeneous treatment of a large family of cluster physical parameters  obtained from their variable star population, e.g. \citet{Arellano2022} and the many works cited there. In the present paper we perform such analysis. The paper is organized as follows: Observations, data reduction and transformations to the standard system are described in \S 2. In \S 3 we describe the variable stars in the field of the cluster including the already known, newly discovered in the present paper and those reported by $Gaia$. The membership analysis of the variable stars is described in \S 4. A differential reddening analysis in the field of the clusters is given in \S 5. The cluster mean metallicity and distance obtained from the Fourier decomposition of RR Lyrae stars are reported in the \S 6. In \S 7 we summarize our results. In Appendix A we comment on peculiar stars.

\section{Observations and Reductions}
\label{sec:Observations}

\subsection{Observations}

The observations were performed from two sites, the 1.54 m telescope of the Bosque Alegre Astrophysical Station of the Córdoba Observatory, National University of Córdoba, Argentina (EABA), and with the 1.0-metre Swope Telescope of the Las Campanas Observatory, Chile (LCO). Data were collected in several dates between 2017 and 2019 as follows: 1) Five nights of 2017 between June 3 and July 29 and three nights of 2018 between May 18 and June 24 in EABA (EABA 17-18); a couple of hours on 28 June 2018 in Swope Telescope of LCO (SWOPE 18); and finally nine nights of 2019 between April 6 and June 30 in EABA (EABA 19). In EABA the camera employed in 2017 and 2018 was Alta F16M with a detector KAF-16803 of 4096 × 4096 square 9-micron pixels, binned 2 × 2, with a scale of 0.496 arcsec/pix after binning, the field of view (FoV) is 16.9x16.9 arcmin$^2$; in 2019 the camera was Alta U9 with a detector KAF-6303E of 3072 × 2048 square 9-micron pixels, binned 2 × 2, with a scale of 0.496 arcsec/pix after binning, the FoV is 12.7 × 8.5 arcmin$^2$. The CCD in LCO was E2V 231-84 of 4096 × 4112 square 15-micron pixels also binned 2 × 2; the scale is 0.435 arcsec/pixel after binning, and the FoV is 14.8 × 14.9 arcmin$^2$.

\begin{table}
\footnotesize
\caption{The distribution of observations of NGC 6139.$^{*}$}
\centering
\begin{tabular}{@{}lccccc}
\hline
Date  &  $N_{V}$ & $t_{V}$ (s) & $N_{I}$ &$t_{I}$ (s)&Avg seeing (") \\
\hline
20170604 & 27 & 400 & 23 & 200-300 & 3.2 \\
20170617 & 21 & 400 & 21 &   200   & 5.6 \\
20170701 & 29 & 400 & 26 &   200   & 2.8 \\
20170702 & 21 & 400 & 22 &   200   & 2.9 \\
20170729 &  6 & 400 & 10 &   200   & 2.5 \\
20180519 & 19 & 400 & 24 &   200   & 3.2 \\
20180520 & 13 & 400 & 16 &   200   & 3.4 \\
20180624 &  1 & 400 &  7 &   200   & 4.0 \\
20180628 &  8 & 10-200 &  8 & 15-60  & 1.8 \\
20190407 & 25 & 300 & 34 & 150 & 2.3 \\
20190408 & 31 & 300 & 35 & 150 & 2.5 \\
20190412 & 30 & 150 & 34 & 300 & 2.9 \\
20190413 & 28 & 300 & 45 & 150 & 3.0 \\
20191414 & 21 & 300 & 24 & 150 & 2.8 \\
20190429 & 17 & 300 & 33 & 150 & 3.4 \\
20190511 & 34 & 300 & 43 & 150 & 3.0 \\
20190629 & 15 & 300 & 24 & 150 & 2.8 \\
20190630 & 14 & 300 & 15 & 150 & 2.6 \\

\hline
Total:   & 360 &        & 444 &           &\\
\hline
\end{tabular}
\center{*: Columns $N_{V}$ and $N_{I}$ give the number of images taken with the $V$ and $I$ filters respectively. Columns $t_{V}$ and $t_{I}$ contain the exposure time. The average seeing is listed in the last column.}
\label{tab:observations}
\end{table}

A total of 360 and 444 images were obtained in the Johnson-Kron-Cousins $V$ and $I$ filters, respectively. 

A detailed log of observations is given in Table \ref{tab:observations} where also the average nightly seeing is recorded.

\subsection{Difference Image Analysis}
We employed the technique of difference image analysis (DIA) to carry out high-precision photometry for all of the point sources in the images of NGC 6139. The {\tt DanDIA}\footnote{DanDIA is built from the DanIDL library of IDL routines available at http://www.danidl.co.uk} pipeline was employed for the data reduction process \citep{Bramich2008,Bramich2013}. Reference images are built for the $V$ and $I$ filter by stacking the best-quality images in each collection. Individual images in each filter are then subtracted from the respective convolved reference image. Differential fluxes for each star detected in the reference image were then measured on each difference image. The total flux per epoch $t$ per star in ADU/s is calculated as:
\begin{equation}
f_{\mbox{\scriptsize tot}}(t) = f_{\mbox{\scriptsize ref}} +
\frac{f_{\mbox{\scriptsize diff}}(t)}{p(t)},
\label{eqn:totflux}
\end{equation}
where $f_{\mbox{\scriptsize ref}}$ is the reference flux,
$f_{\mbox{\scriptsize diff}}(t)$ is the differential flux and
$p(t)$ is the photometric scale factor (the integral of the kernel solution).
Conversion to instrumental magnitudes is done via:
\begin{equation}
m_{\mbox{\scriptsize ins}}(t) = 25.0 - 2.5 \log \left[ f_{\mbox{\scriptsize tot}}(t)
\right],
\label{eqn:mag}
\end{equation}
where $m_{\mbox{\scriptsize ins}}(t)$ is the instrumental magnitude of the star 
at time $t$. 
  The above procedure is described in further detail in \cite{Bramich2011}.

\subsection{Transformation to the Standard System}

Standard stars in the field of NGC 6139 are found in the online collection of \cite{Stetson2000} \footnote{http://www3.cadc-ccda.hia-iha.nrc-cnrc.gc.ca/community/STETSON/standards}. We use 115 stars with $V$ and $I$ standard values in the FoV of our images, to transform instrumental $vi$ magnitudes into the standard \emph{VI} system.

\begin{table}
\footnotesize
\caption{Time-series $V$ and $I$ photometry for all the variables in the field of view of NGC 6139. (Full table is available in electronic format).}
\centering
\begin{tabular}{ccccc}
\hline
Variable &Filter & HJD & $M_{\mbox{\scriptsize std}}$ &
$\sigma_{m}$  \\
Star ID  &        & (d) & (mag) &(mag)\\
\hline
V2& V & 2457908.50673&14.870& 0.004\\
V2& V & 2457908.70500&14.876& 0.003\\
\vdots   & \vdots & \vdots  & \vdots & \vdots  \\
V2& I & 2457908.49728&12.324& 0.002\\
V2& I & 2457908.50246&12.320& 0.002 \\
\vdots   & \vdots & \vdots  & \vdots & \vdots  \\
V5& V & 2457908.50673&17.164& 0.023\\
V5& V & 2457908.53086&17.123& 0.018\\
\vdots   & \vdots & \vdots  & \vdots & \vdots \\
V5& I & 2457908.49728&16.018& 0.024\\
V5& I & 2457908.50246&16.005& 0.022\\
\vdots   & \vdots & \vdots  & \vdots & \vdots \\
\hline
\end{tabular}
\label{tab:vi_phot}
\end{table}

\begin{figure*}
    \centering
    \includegraphics[width=17.0cm]{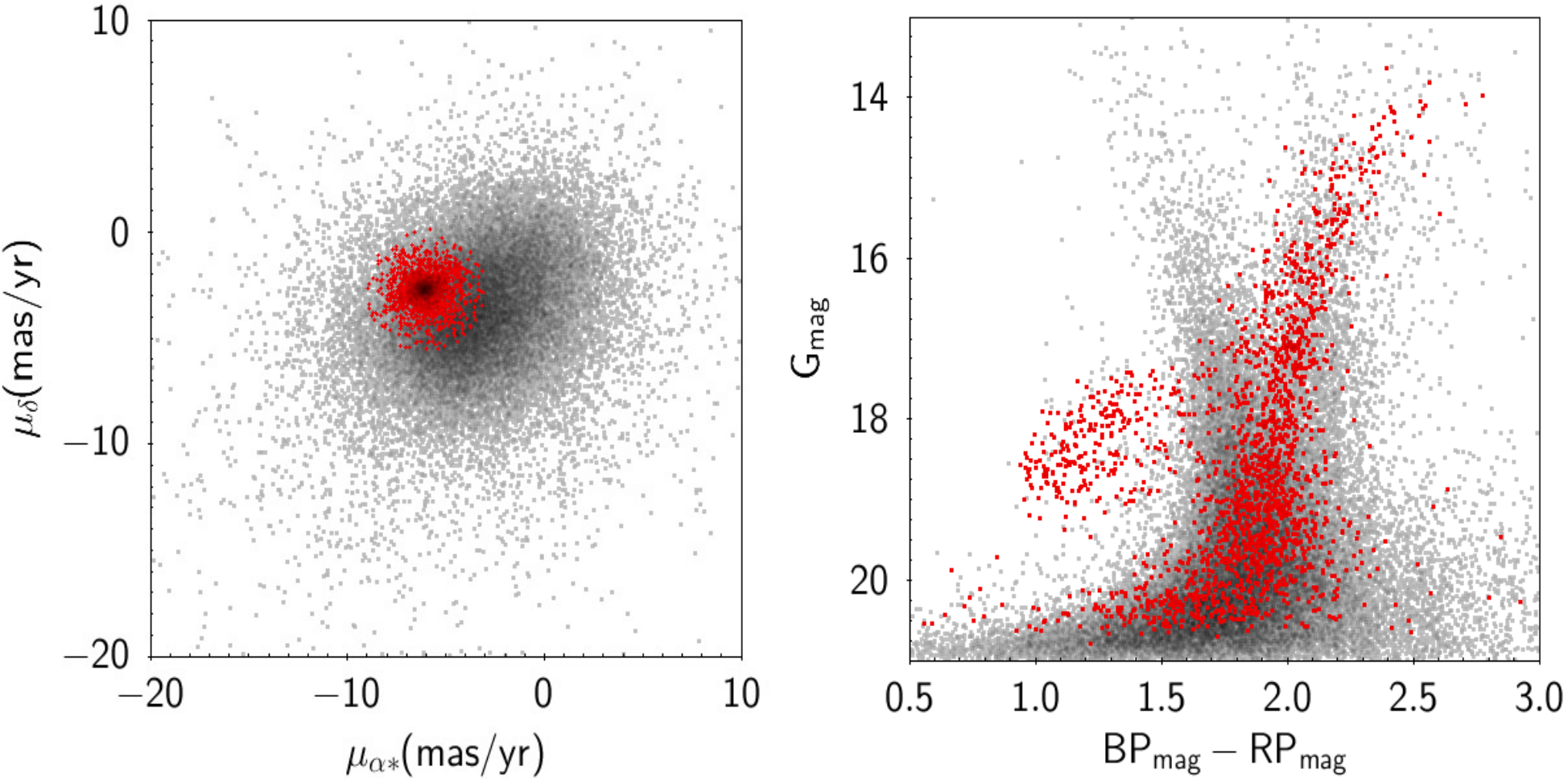}
    \caption{VPD and CMD of field (gray dots) and member stars (red dots) of NGC\,6139. }
    \label{VPD-CMD}
\end{figure*}

The transformation equations between the instrumental and the standard magnitudes were independently calculated for observations from different detectors and observatories. The transformation equations are of the form $V-v = A (v-i) + B$, and $I-i= C (v-i) + D$, where capital $V$ and $I$ refer to the standard system and lower case  $v$ and $i$, to the instrumental one.In Table \ref{tab:coeff} we include the values of coefficients for each observing season. For an explicit example of the transformation plots see for instance  \citet{Yepez2022}.

\begin{table}
\footnotesize
\caption{Season transformation coefficients for equation of the form:$V-v = A (v-i) + B$, and $I-i= C (v-i) + D$. }
\centering
\begin{tabular}{cccc}
\hline
Coeff. & EABA 17-18    &     SWOPE 18       &    EABA 19 \\
\hline
A &  0.032 $\pm$ 0.011 & -0.054 $\pm$ 0.009 &  0.086 $\pm$ 0.017 \\
B & -2.871 $\pm$ 0.011 & -1.510 $\pm$ 0.020 & -2.865 $\pm$ 0.023 \\
C &  0.017 $\pm$ 0.011 &  0.078 $\pm$ 0.004 &  0.047 $\pm$ 0.016 \\
D & -3.843 $\pm$ 0.010 & -1.478 $\pm$ 0.010 & -3.420 $\pm$ 0.021 \\
\hline
\end{tabular}
\label{tab:coeff}
\end{table}

All of our \emph{VI} photometry for the studied variables in this work is provided in Table \ref{tab:vi_phot}. A small portion of this table is given in the printed version of this paper and the full table is available in electronic form in the {\it Centre de Donnes astronomique de Strasbourg} data base (CDS). 

\section{Cluster star membership and the variables in the color-magnitud diagram.}

The membership status of all stellar sources in the FoV of our images was investigated based on the astrometric data available in {\it Gaia}-DR3 \citep{Gaia2016,Gaia2022} using the method developed by \cite{Bustos2019}. Such method consists of two stages; 1) it aims to find groups of stars that possess similar characteristics in the four-dimensional space of the gnomonic coordinates ($X_{\rm t}$,$Y_{\rm t}$) and proper motions ($\mu_{\alpha*}$,$\mu_\delta$) by means of a clustering algorithm and 2) the analysis of the projected distribution of stars with different proper motions around of the mean proper motion of the cluster, in order to extract likely members that were missed in the first stage. The specific details about this method are given in the paper quoted above. 

The method was applied to 45717 $Gaia$-sources located within a radius of 15 arcmin from the cluster center, 30412 of which with proper motions in $Gaia$-DR3. This selected radius was chosen to guarantee  a proper sampling of the cluster background. Only 2339 were found to be cluster members; our differential photometry was able to produce light curves for 1108 of these star.

The vector point diagram (VPD) and CMD of field and member stars are shown in Fig. \ref{VPD-CMD}.

\section{Variable stars in NGC 6139}

The Catalogue of Variable Stars in Globular Clusters (CVSGC) \citep{cle01}, in its February 2013 edition, lists 10 variables in NGC 6139, mostly asymetrically distributed towards the cluster peripheria. These variables were all discovered in photographic plates by Hazen (1991) and no other time-series study of the cluster is available after it. Intuition makes us believe that there must be more, not yet discovered, variables in the field of the cluster, some of which might be cluster members.

\begin{figure*}
\includegraphics[width=17.0cm]{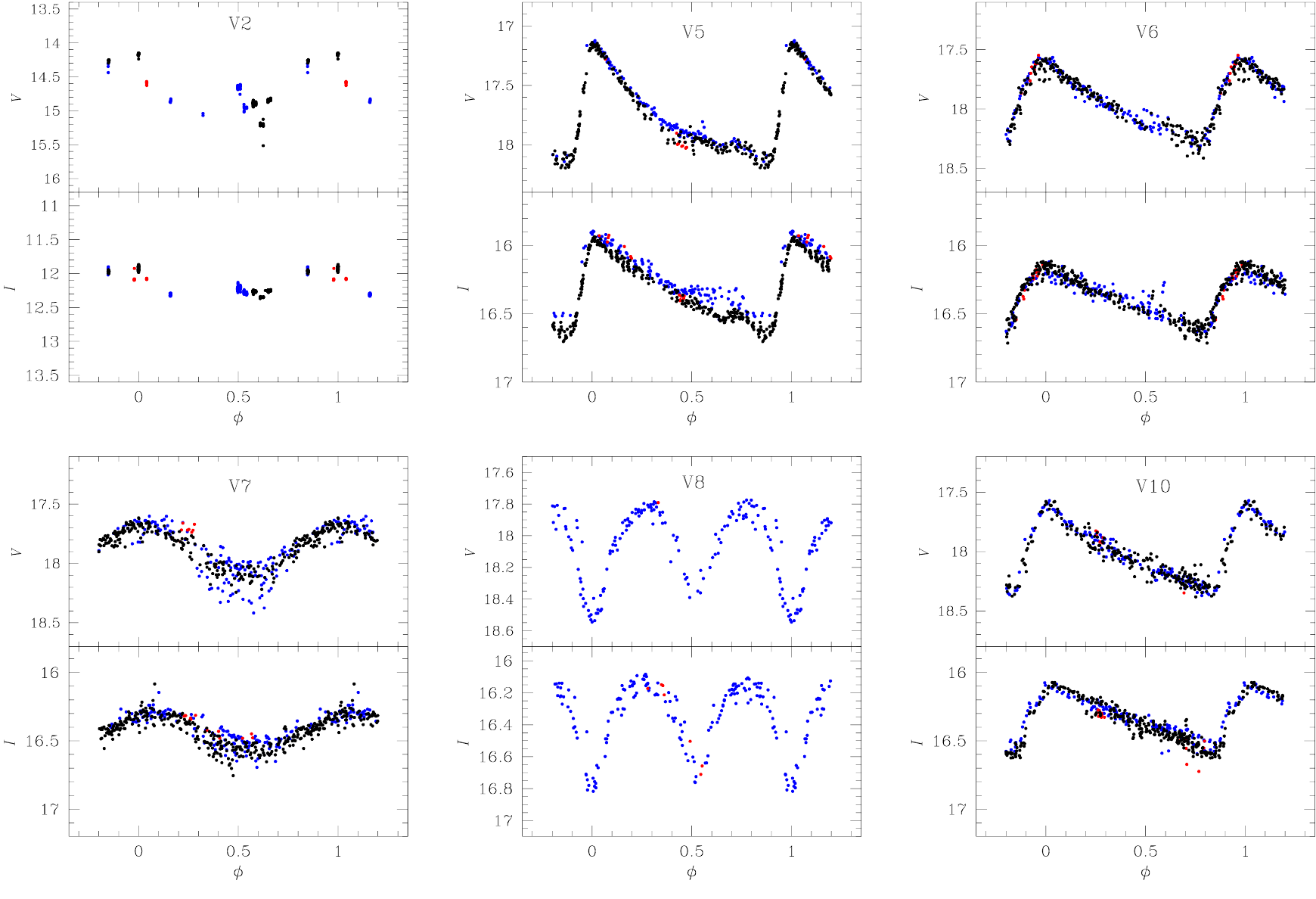}
\caption{Light curves of known variables previous to this work in the FoV of our images, phased with the periods listed in Table \ref{variables}. Colour symbols are blue for Bosque Alegre 2017-2018 data, red for 2018 from Las Campanas SWOPE and black for and Bosque Alegre 2019 data.}
\label{knownvar}
\end{figure*}

Only six of the ten variables listed in the CVSGC are found within the FoV of our images; V2 (long period variable, L), V5, V6 and V10 (RRab),  V7(RRc) and V8 (eclipsing binary, EB). The light curves of these stars in our data are shown in Fig. \ref{knownvar} and are phased with the ephemerides given in Table \ref{variables}.

\begin{table*}
\small
\caption{General data for the variable stars in the FoV of NGC 6139.}
\label{variables}
\centering
\begin{tabular}{clccccccccc}
\hline
Variable & Variable & $<V>$$^1$    & $<I>$$^1$   & $A_V$       & $A_I$   & $P$ (days)    &
 HJD$_{\rm max}$     &  RA          & Dec.& member (m)/ \\
Star ID  & Type$^{a}$& (mag)   & (mag)   & (mag)       & (mag)   & this work     
&  (d +245~0000.)    &  (J2000.0)   & (J2000.0)& field (f) \\
\hline
V2 & SR   & 14.647 &12.176  & 1.04 & 0.40 & 79.87 &8614.7526 &16:27:12.54 & -38:49:28.9& m \\
V5 & RRab & 17.777 & 16.316 & 0.893 & 0.707 & 0.594961 & 8663.5174 &16:27:34.31 & -38:51:19.3& m \\
V6 & RRab & 17.975 & 16.391 & 0.592 & 0.468 & 0.705903 & 8614.7690 & 16:27:43.79 & -38:48:21.7& m \\
V7 & RRc  &17.901 & 16.436 & 0.372& 0.301 & 0.420590 & 8586.9055 & 16:27:44.53 & -38:47:59.5& m \\ 
V8 & EC   & 18.065 &  16.352 & 0.724 & 0.707 & 0.353222 & 7936.6251 & 16:28:16.40 & -38:49:58.9& f \\
V10& RRab & 18.015 & 16.338 &0.613& 0.514& 0.758156& 8581.8286& 16:27:26.32 & -38:49:07.3& m \\
\hline
V11 (G4)$^3$& RRc  & 17.920 & 16.711&0.453& 0.293& 0.318837& 8614.5774& 16:27:38.32 & -38:53:10.6& m\\
V12$^2$ & RRc  & 17.823 & 16.613 & 0.453 & 0.293 & 0.347322 & 8298.4985 & 16:27:37.56 & -38:52:07.5& m \\
V13$^2$ & SR   & 14.873& 12.213 & 0.482 & 0.345 & 189.0588 & 8664.6697 & 16:27:40.02 & -38:50:56.9& m \\
V14$^2$ & SR   & \it 14.705& \it 12.273  & 0.482 & 0.345 &    --    &    --     & 16:27:40.56 & -38:51:05.0& m \\
V15$^2$ & RRc? & 17.819 & 15.928 & 0.214 & 0.125 & 0.364931 & 8664.6518 & 16:27:42.24 & -38:50:59.8& m \\
V16 (G3)$^3$& RRc  & 17.719 & 16.564 & 0.228& 0.155 &0.321500& 7908.7357& 16:27:35.56 & -38:51:42.6& m\\
V17$^2$ & SR & 15.196 & 12.777 & 0.134 & 0.114 & 27.5010 & 8614.7329 & 16:27:32.88 & -38:50:42.2& m \\
V18$^2$ & SR & 15.725 & 12.674 & 0.254 & 0.200 & 85.1571 & 8257.6707 & 16:27:41.77 & -38:50:33.0& m \\
\hline
G1$^3$& EB  & \it 18.586 & \it 17.169 & 0.747& 0.359 &0.412506& 8602.8116& 16:27:15.27 & -38:54:13.6& f \\
G2$^3$& SR  & 17.314 & 13.184 & 0.345& 0.205 &102.89& 8298.4741& 16:27:23.54 & -38:53:40.1& f \\
G5$^3$& EB  & 18.471 & 16.716 & 0.506& 0.468 &0.374929& 8585.8395& 16:27:47.91 & -38:49:44.5& f\\
G6$^3$& SR  & 17.692 & 12.853 &0.79& 0.57 &--&--& 16:27:55.29 & -38:54:54.5& f\\
G7$^3$& EB  & 18.441 & 17.067 & 0.338& 0.330 &0.480836& 8585.6825 & 16:27:58.23 & -38:50:54.0& f\\
G8$^3$& EB  & 17.885 & 16.239 & 0.177& 0.148 &0.40324& 8586.8631 & 16:28:11.19 & -38:48:24.1& f\\
G9$^3$& EB  & 18.470 & 16.417 & 0.521& 0.329 &0.864273& 7908.5309 & 16:28:09.27 & -38:45:19.9& f\\
G10$^3$& SR  & \it 16.096 &12.788 & 0.189& 0.099 &30.0& 7936.7083 & 16:27:59.43 & -38:56:20.7& f\\
\hline
N1$^2$  & RRc/EC? & 18.235 & 16.817 & 0.167 & 0.133 & 0.537526 & 8586.9126 & 16:27:40.55 & -38:49:57.2& f \\
    &         &    &    &    &    & 0.268666 &           &              & &         \\
N2$^2$  & RRab?  & 16.321 & \it 14.004 & 0.034 & -- & 0.796277 & 7921.7212 & 16:27:17.48 & -38:55:23.1 & f\\
N3$^2$  & RRc & 16.991 & 15.572 & 0.19 & 0.20 & 0.342243 & 7963.6339 & 16:28:13.23 & -38:52:49.0 & f\\
N4$^2$ & SR? & 14.869 & 11.514 & 0.342 & 0.136 & 24.7644 & 7921.7817 & 16:27:25.45 & -38:48:42.7& f \\

\hline
\end{tabular}

\center{1: Intensity weighted means unless in italics, in which case magnitudes are simple means.\\

2: Variable newly found in this paper.\\

3: Variable found by $Gaia$-DR3, confirmed, classified and ephemeris from our data in present work.\\
}
\end{table*}

\begin{figure*}
\includegraphics[width=17.0cm]{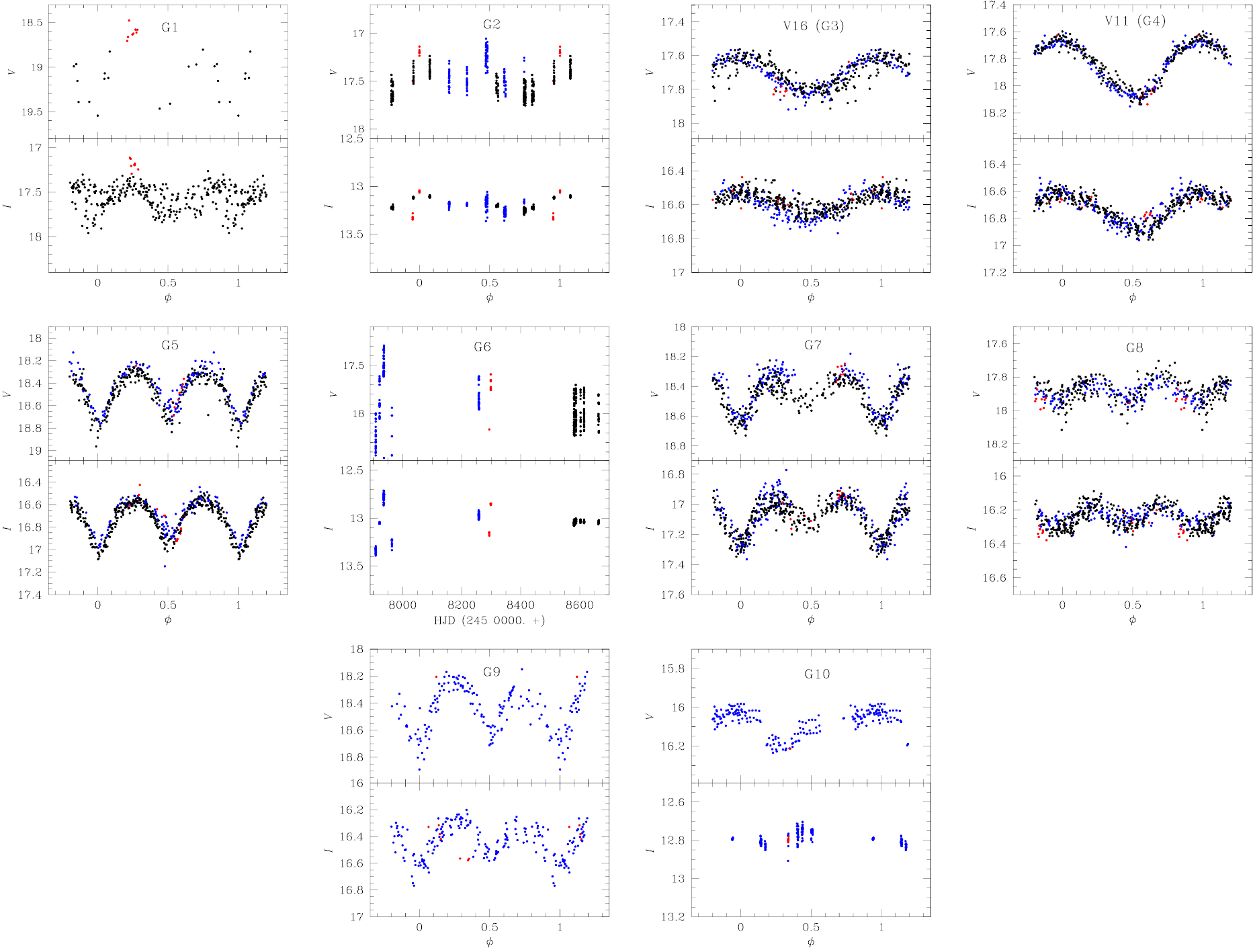}
\caption{Light curves variables included as such in $Gaia$-DR3 and confirmed in our data collection. The assigned variable type and ephemerides are listed in  Table \ref{variables}. Colour symbols are as in Fig. \ref{knownvar}.}
\label{GaiaVar}
\end{figure*}

Most recently, in the data base of $Gaia$-DR3 \citep{Gaia2016,Gaia2022}, there are reported 86 variables of assorted types in the grand field of NGC 6139, which includes the 10 known variables in the CVSGC. Only 18 of the non previously known variables are within the FoV of our images. Hence we are in position to corroborate their variability, calculate their periodicity, rectify their classification and argue on their cluster membership.  We confirmed the variability of 10 of theses stars and called them G1 to G10. Their light curves are shown in Fig \ref{GaiaVar}. Not all of these stars seem to pertain to the cluster. According to our membership analysis, only G3 and G4 seem to be cluster members, thus we assigned them new variable names V16 and V11 respectively.

\begin{figure*}
\centering
\includegraphics[width=17.0cm]{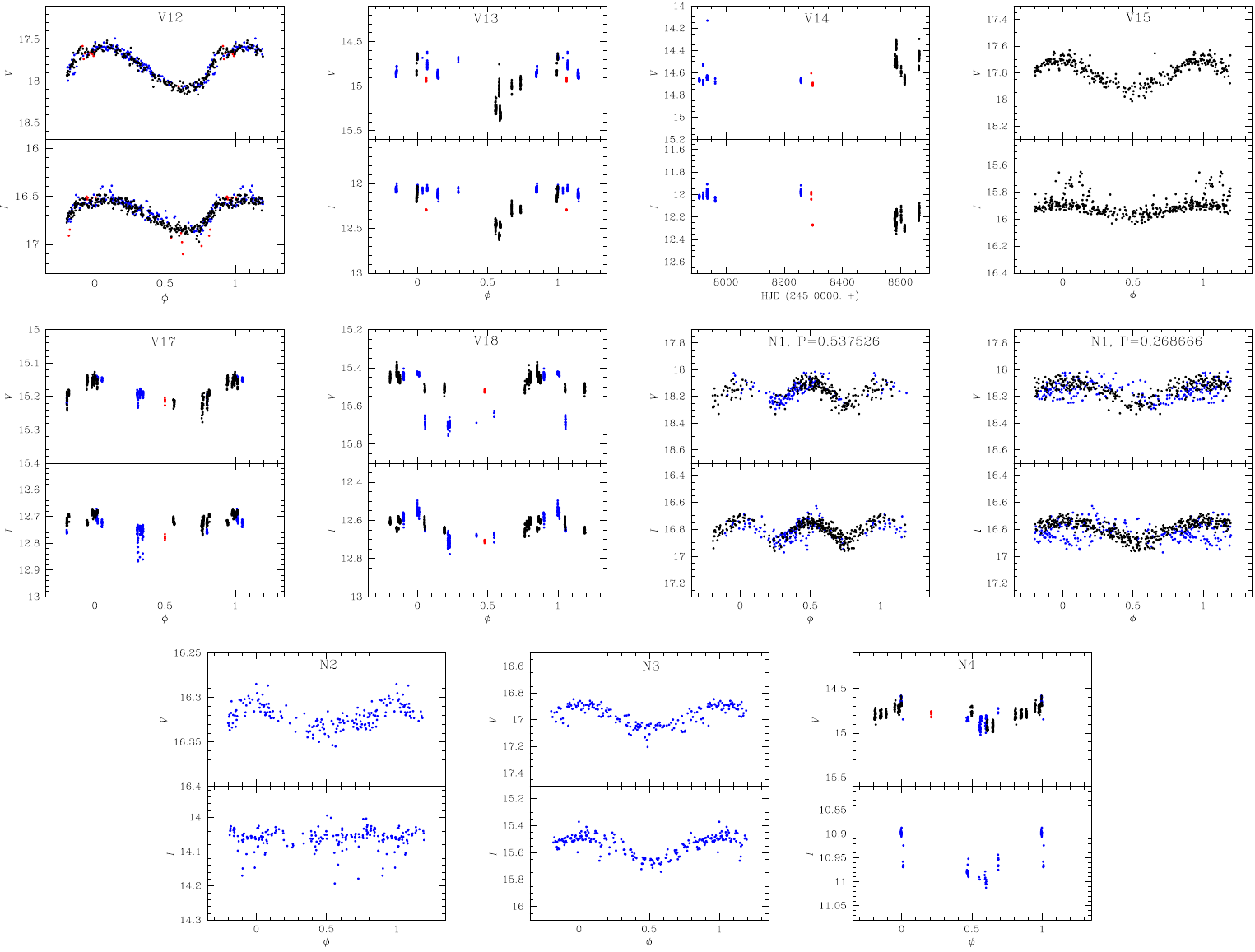}
\caption{Light curves of new variables discovered in this work in the FoV of our images. The assigned variable type and ephemerides are listed in  Table \ref{variables}. Colour symbols are as in Fig. \ref{knownvar}. The variable N1 is displayed phased with the two competing periods found int Table \ref{variables}. The variability of V14 is marginal and should be confirmed with further observations. The periodicity of N4 will likely be refined from further observations and a longer time span. See comments in Appendix \ref{sec:comments} }.
\label{newvar}
\end{figure*}

Taking advantage of our dense time-series \emph{VI} CCD photometry, we have searched for other variables and found 10 not previously detected variables. Six of these variables are likely members, hence we identified them as V12-V15 and V17-V18. We named the remaining four like N1 to N4.

The light curves of all newly discovered variables are displayed in Fig. \ref{newvar}.

Therefore, Table \ref{variables} is organized in four groups as follows; 1) previously known variables, 2) newly discovered variables and those from the $Gaia$-DR3 source confirmed as variables that proof to be likely cluster members, and for which we assigned avariable nomenclature (V11 to V18), 3) confirmed $Gaia$-DR3 variables, with new estimations of their period and a classification. No variable name is assigned to these variables since they are found to be field stars. And 4) newly discovered variables in our data but apparently no members of the cluster. Hence, the $V$ nomenclature has been reserved for cluster members. G is used for field variables listed by $Gaia$-DR3 and N for field newly found variables in the present work. Table \ref{variables} also contains intensity weighted means $<V>$, $<I>$, amplitudes and coordinates taken from the $Gaia$-DR3. All variables in Table \ref{variables} are identified in the charts of Fig. \ref{CHARTS}, and detailed comments on peculiar specific variables are given in Appendix \ref{sec:comments}.

\begin{figure*}
\centering
\includegraphics[width=16.0cm]{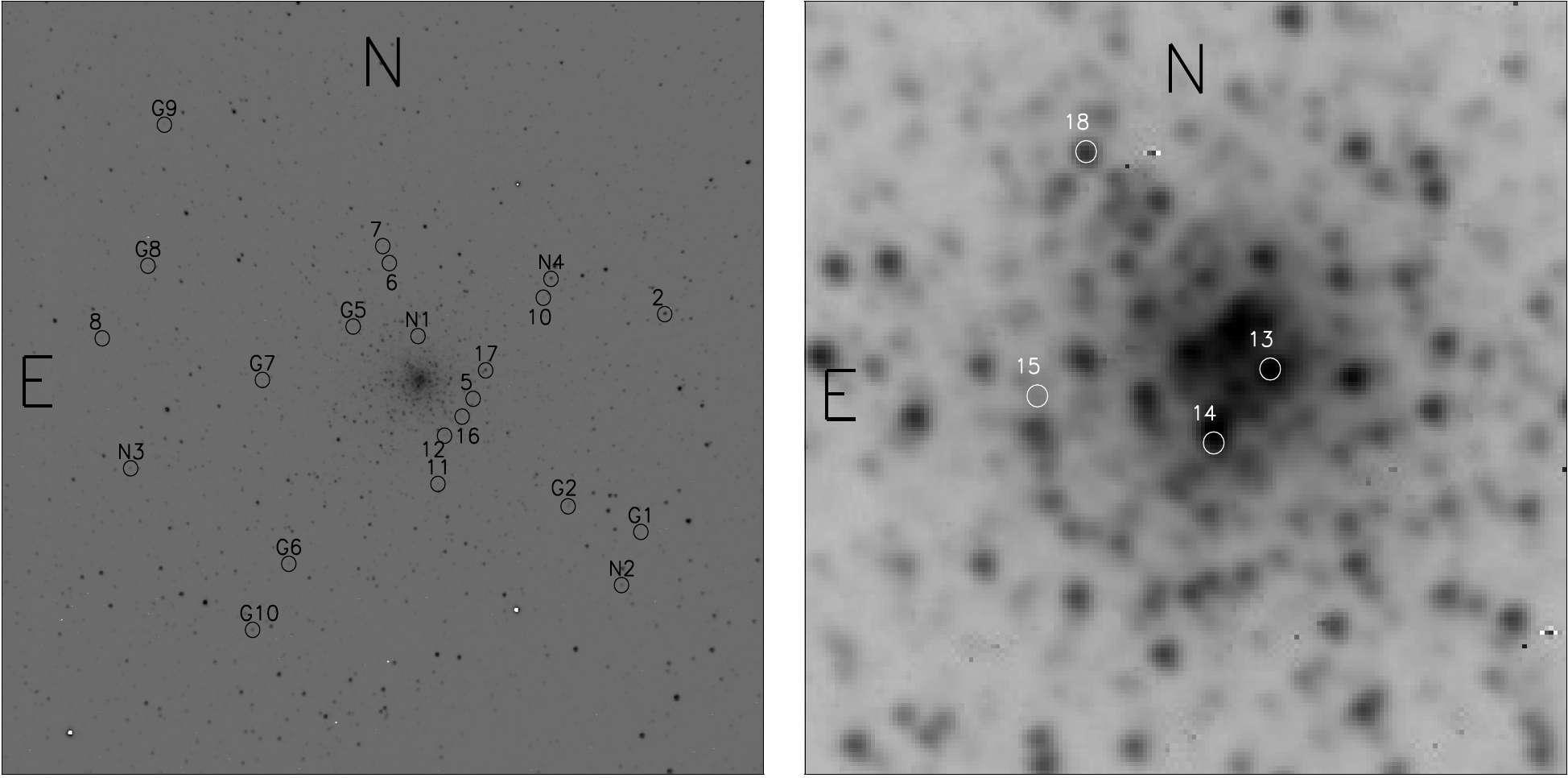}
\caption{Identification charts of the confirmed variables in NGC 6139 (see Table \ref{variables}). The left panel shows the FoV of our images from EABA 2017-2018, the field size is about 16.8$\times16.8$ arcmin$^{2}$. The panel on the right displays only the central region of the cluster and the field is of 1.4$\times1.4$ arcmin$^{2}$.}
\label{CHARTS}
\end{figure*}

\subsection{Periods}
\label{sec:Periods}

The periods of all variables were calculated using the string length method \citep{Burke1970, Dworetsky1983} which searches for the period that phases the light curve with a minimum dispersion. For the long period variable, whose light curves are not necessarily fully covered, a first approach was attempted with {\tt period04} \citep{Lenz2005} that calculates a discrete Fourier transformation and calculates a least-squares fitting of multiple frequencies to the data and an associated amplitude-frequency spectrum. Then the periods were refined via the string length method. The final periods are listed in Table \ref{variables} and where used to phase all light curves displayed in the figures.

\subsection{The Oosterhoff type of NGC 6139}

\begin{figure}
\includegraphics[width=8.0cm]{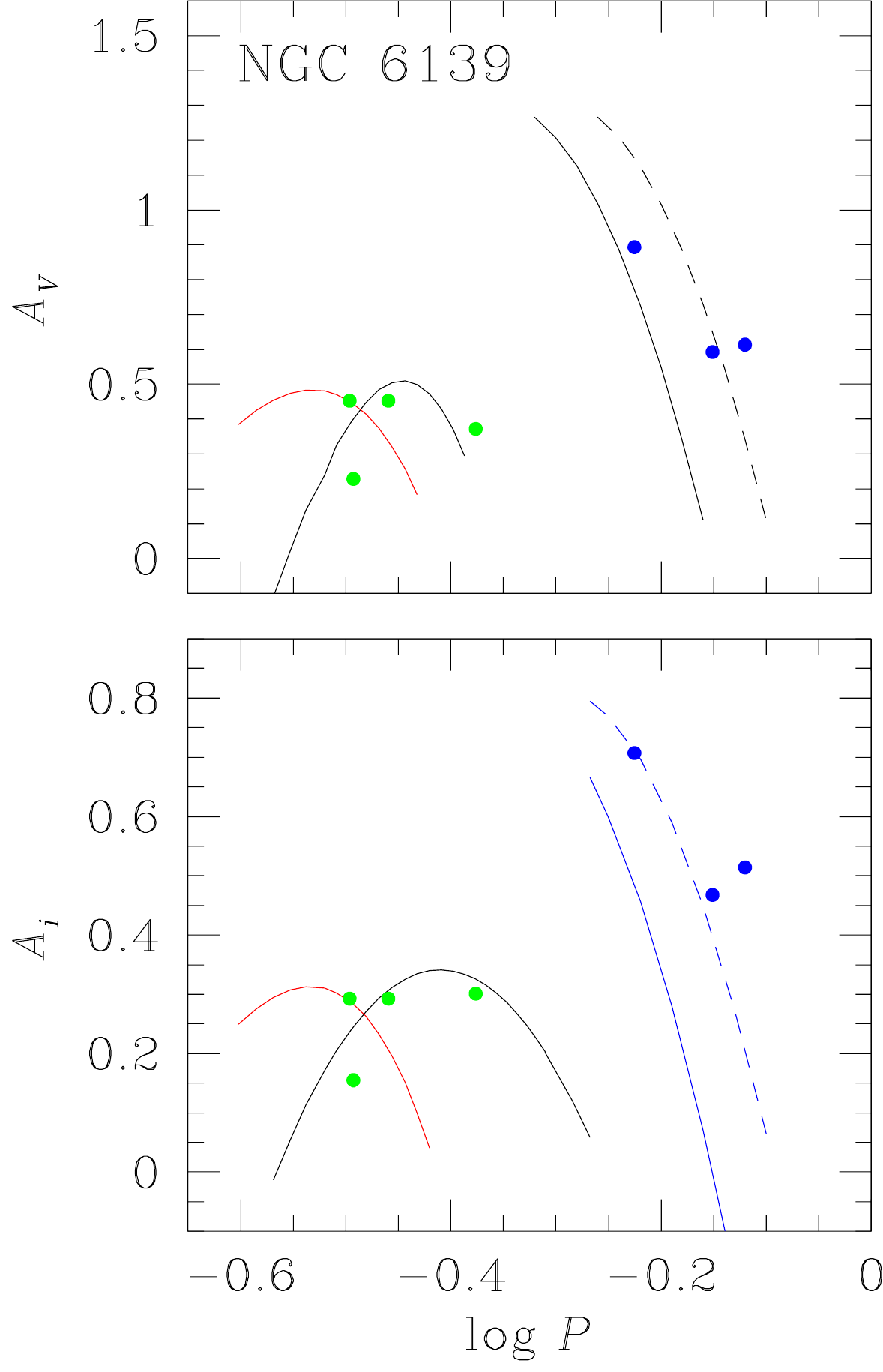}
\caption{Period-Amplitude diagram for RR Lyrae stars in NGC 6139. Blue and green symbols represent RRab and RRc stars respectively. The curves in the top panel and to the right, are the locy for RRab stars (unevolved continuous and evolved segmented) in M3 according to \citep{Cacciari2005}. \citet{Kunder2013b} found the black parabola for the RRc stars from 14 OoII clusters and \citet{Arellano2015} calculated the red parabolas from a sample of RRc stars in five OoI clusters, excluding variables with Blazhko effect. In the bottom panel, the continuous and segmented blue lines were constructed by \citet{Kunder2013a}. The  black parabola was calculated by \citet{Yepez20}, using 28 RRc stars from seven OoII clusters.}
\label{Bailey}
\end{figure}

Oosterhoff types for clusters with few RR Lyraes are not easy to determine using the average period of their RRab stars or the RRc/RRab ratios. Also when metallicities are in the mid range of about -1.5 dex, the overlap between Oo I and Oo II clusters is substantial, hence, metallicity alone is not a truthful discriminator. In these cases the Amplitude-Period diagram, or Bailey's diagram, is a better indicator. In the case of NGC 6139 we have three RRab and four RRc stars. Their amplitudes and periods are plotted in the Bailey's diagram of Fig. \ref{Bailey}. It is clear that both RRab and RRc stars are distributed towards the evolved sequences, which identifies the cluster as of the Oo II type.

\section{Differential Reddening}
\label{sec:redd} 
NGC 6139 is located near the Galactic Bulge and given the environment in such region, it may be subject to interstellar differential reddening. To our knowledge, no differential reddening map is  available.

For RRab stars an independent approach to their reddenings is by recalling that intrinsic colour $(B-V)_0$ at minimum light in nearly constant \citep{Sturch1966} and that $(V-I)_{0;min}$ = 0.58$\pm$0.02 mag \citep{Guldenschuh2005}. The method, applied to the three cluster member RRab stars, V5, V6 and V10, gave an average of  $0.828 \pm 0.038$ mag where the uncertainty is the standard error of the mean. This value could be compared with results for $E(B-V)$ of $0.786 \pm  0.016$ and  $0.913 \pm 0.019$ mag from the interstellar extinction calibrations of \cite{Schlafly2011} and \citet{Schlegel1998} respectively. A revision of the dereddened CMD showed that for $E(B-V)=$0.786 mag we obtained the best agreement between the observations and the theoretical loci of the HB and Red Giant Branch (RGB), and then this value was adopted as the mean reddening for the differential analysis described below. 

\begin{figure}
    \centering
\includegraphics[width=7.7cm]{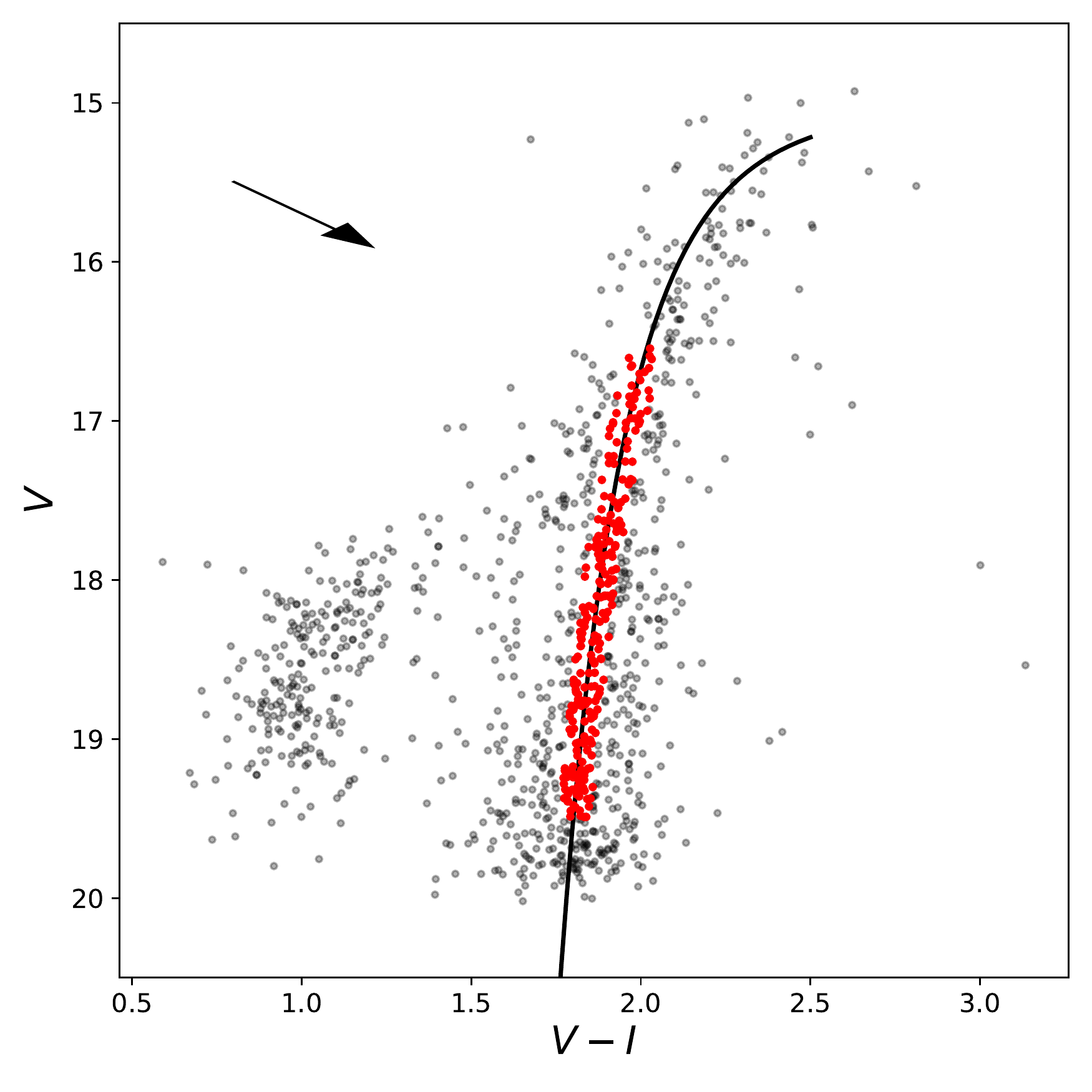}
    \caption{Colour-magnitude diagram of NGC\,6139. The solid line is the mean ridge line (MRL). The stars used as reference to derive $\delta E(B-V)$ are shown as red points. The reddening vector was calculated for an extinction $A_V=3.12~E(B-V)$, $E(V-I)/E(B-V)=1.259$ and $E(B-V)= 0.756$.}
    \label{fig:cmd_mrl}
\end{figure}

\begin{figure}
    \centering
    \includegraphics[width=8.0cm]{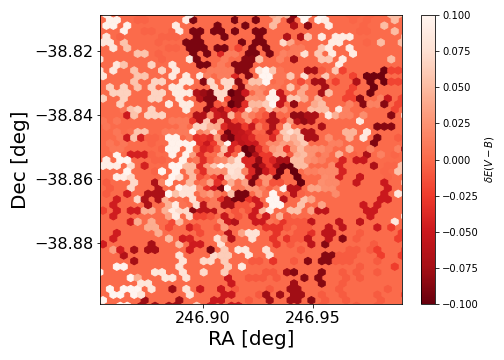}
    \caption{Differential reddening map. Each cell has a resolution of $ 7'' \times 8''$. The values of $\delta E(B-V)$ in each cell are the mean for the stars within such region.}
    \label{fig:difred_map}
\end{figure}

\begin{figure}
    \centering
    \includegraphics[width=8cm]{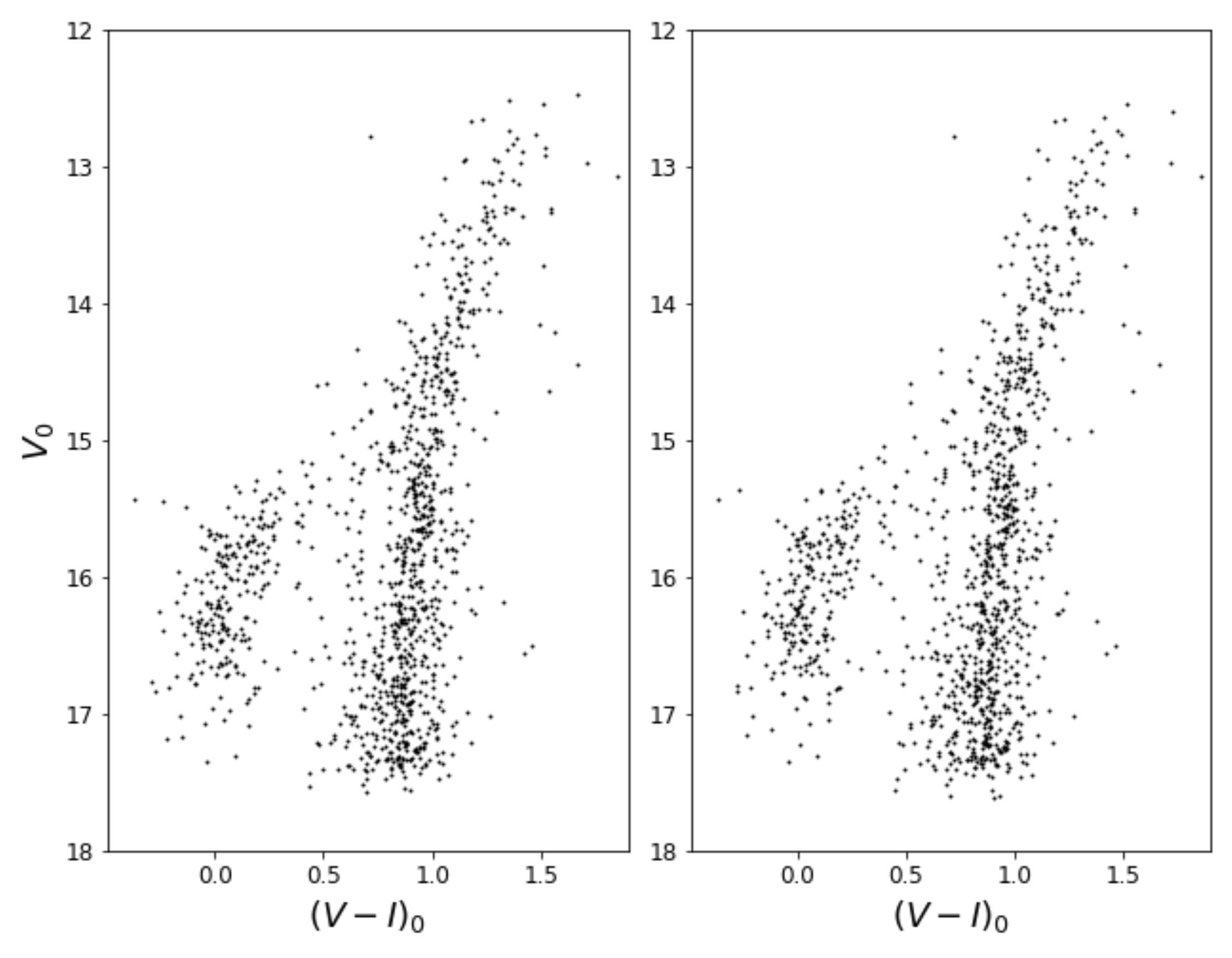}
    \caption{Colour-magnitude diagram corrected by reddening (left) and corrected also for differential reddening (right, after adding an offset in $(V-I)_0$).}
    \label{fig:cmd_difred}
\end{figure}

However, we have attempted the construction of a differential reddening map for NGC 6139 and proceeded as follows. Along the RGB in the CMD, within $16.5 \leq V \leq 19.5$, the photometric errors are precise enough so that the scatter in the $V-I$ colour is at least partially due to differential reddening. We have splitted the above magnitude range into 0.1 mag bins, and calculated the mean $V-I$ of the stars in each bin. Using a cubic spline to interpolate such points, we compute the mean ridge line (MRL), shown in Fig. \ref{fig:cmd_mrl}.

We select as reference stars, those within 2$\sigma$ of the MRL along the RGB within the $V$ magnitude range described above. Such stars are shown in Fig. \ref{fig:cmd_mrl} as red points. Then, we compute the distance $\Delta X$ to the MRL along the reddening vector shown in the figure, and recalling that assuming an average $E(B-V)=0.786$ mag, $A_V= 3.12 \times 0.786 = 2.45$ mag.

The differential reddening is obtained form $\Delta X$ as follows:
\begin{equation}
    \delta E(B-V)= \frac{\Delta X}{R_V-R_I}
\end{equation}

\noindent
with $R_V=3.120 pm$0.002 and $R_I=1.90$ $\pm$0.02 \citep{Casagrande2014} which translates into an average uncertainty in $\delta E(B-V)$ of $\pm 0.0008$ mag.

This method gives an individual value of $\delta E(B-V)$ for each reference star \citep[see e.g.,][]{Deras2023,Saracino2019}. We associate a $\delta E(B-V)$ value to each cluster member star by interpolation with the nearest reference stars. A plot of the differential reddening for the cluster member stars in the RA-Dec plane leads to the differential reddening map of Fig \ref{fig:difred_map}, where each cell has a size of 7” x 8” and its colour depends on the mean $\delta E(B-V)$ of the stars within the cell. The differential reddening ranges between $-0.1<\delta E(B-V)<0.1$.

Figure \ref{fig:cmd_difred} shows the dereddened CMD before and after applying the differential reddening correction. The scatter is lower after the differential reddening correction, but not substantial. This, in spite of the low uncertainty in $\delta E(B-V)$, may be due to the following reasons; the low number of reference stars available for this analysis, and the fact that, although differential reddening in the grand field of the cluster may be substantial, when our analysis is concentrated closer to the cluster field the differential reddening is  milder (see for example the effect in figure. 5 of \citet{Ortolani1999}). Although there are archival HST observations, that produce a deeper CMD (two magnitudes bellow the MSTO), such observations only cover the inner part of NGC\,6139, hence not compatible for a differential reddening analysis in our observations.

\section{Light curve Fourier decomposition and physical parameters}
\label{sec:RRLstars}

The approach of RR Lyrae light curve decomposition aimed to estimate physical parameter, such as the metallicity, luminosity, temperature, mass and radii, has been systematically employed in globular clusters, see \citet{Arellano2022} for a detailed description of the technique and empirical calibrations. It has been demonstrated that the metallicity and distance inferred for the RR Lyrae population parental cluster are in excellent agreement with values obtained from high resolution spectroscopic and modern parallax determinations respectively e.g. \citet{Arellano2022} and references there cited. The standard procedure is to represent the $V$ light curves by the Fourier series of harmonics:

\begin{equation}
m(t) = A_0 + \sum_{k=1}^{N}{A_k \cos\left (\frac{2\pi}{P}~k~(t-E_0) + \phi_k \right) },
\label{eq_foufit}
\end{equation}

\noindent
where $m(t)$ is the magnitude at time $t$, $P$ is the period and $E_0$ the epoch, generally a time of maximum light. A
linear minimization routine is used to derive the amplitudes $A_k$ and phases $\phi_k$ of each harmonic, from which the Fourier parameters $\phi_{ij} = j\phi_{i} - i\phi_{j}$ and $R_{ij} = A_{i}/A_{j}$ are calculated. These Fourier parameters and the semi-empirical calibrations of \cite{Jurcsik1996}, for RRab stars, were used to obtain  [Fe/H]$_{\rm ZW}$ on the \cite{Zinn1984} metallicity scale that can be transformed to the UVES scale using the equation [Fe/H]$_{\rm UVES}$=$-0.413$ +0.130~[Fe/H]$_{\rm ZW}-0.356$~[Fe/H]$_{\rm ZW}^2$ \citep{Carretta2009}. The absolute magnitude $M_V$ can be derived from the calibration of \citet{Kovacs2001}  for RRab stars. For the RRc stars the calibration for [Fe/H]$_{\rm ZW}$ of \citet{Morgan2007} was empoyed whereas the absolute magnitude $M_V$ comes from the calibration of \citet{Kovacs1998}.

The corresponding equations towards the calculation of $T_{\rm eff}$, log$(L/{\rm L_{\odot}})$, $M/{\rm M_{\odot}}$ and $R/{\rm R_{\odot}}$ are described in \citet{Arellano2010}.
 
The [Fe/H] calibration for RRab stars of \cite{Jurcsik1996} is applicable to stars with a {\it deviation parameter} $D_m$, defined by these authors, not exceeding an upper limit. This parameter measures how consistent is the morphology of the light curve to evaluate with that of the calibrators employed to defined the [Fe/H] calibration for RRab stars. These authors suggest $D_m \leq 3.0$. The $D_m$ is listed in column 10 of Table~\ref{fourier_coeffs}.

The average  weighted mean [Fe/H]$_{\rm ZW}$ and $M_V$ and the corresponding distances and physical parameters are reported in Table \ref{fisicos}.

\begin{table*}
\footnotesize
\centering 
\caption[]{\small Fourier coefficients for RRab and RRc stars. The numbers in parentheses indicate the uncertainty on the last decimal place. Also listed is the deviation parameter $D_{\mbox{\scriptsize m}}$ for RRab stars.}
\centering                   
\begin{tabular}{lllllllllr}
\hline
Variable ID & $A_{0}$ & $A_{1}$ & $A_{2}$ & $A_{3}$ & $A_{4}$ & $\phi_{21}$ & $\phi_{31}$ & $\phi_{41}$ &  $D_{\mbox{\scriptsize m}}$ \\
   & ($V$ mag)  & ($V$ mag)  &  ($V$ mag) & ($V$ mag)& ($V$ mag) & & & & \\

\hline
  &  &  &  & RRab &  &  &  &  & \\
\hline
V5 & 17.777(2) & 0.336(3) & 0.182(3) & 0.114(3) & 0.083(3) & 3.921(26) & 8.199(40) & 6.118(55) & 2.6\\
V6 & 17.975(3) & 0.261(4) & 0.116(4) & 0.067(4) & 0.024(4) & 4.309(44) & 8.952(70) & 7.529(16) & 4.3\\
V10 & 18.015(3) & 0.255(4) & 0.122(4) & 0.074(4) & 0.040(4) & 4.389(46) & 8.824(71) & 6.920(12) & 2.2\\
\hline
  &  &  &  & RRc &  &  &  &  & \\
\hline
V7 & 17.901 (4) & 0.209(5) & 0.010(5) & 0.004(5) & 0.015(5) & 4.833(502) & 3.614(950) & 3.452(350) & \\
V11 & 17.920(3) & 0.201(4) & 0.022(4) & 0.013(4) & 0.005(4) & 4.827(207) & 4.240(343) & 3.344(897) & \\
V12 & 17.823(2) & 0.249(4) & 0.036(3) & 0.020(3) & 0.015(3) & 4,914(93) & 3.876(165) & 2.418(217) & \\
V16 & 17.719(2) & 0.111(4) & 0.005(4) & 0.003(4) & 0.004(4) & 3.737(816) & 2.916(990) & 1.278(965) & \\
\hline
\end{tabular}
\label{fourier_coeffs}
\end{table*}

\begin{table*}
\footnotesize
\centering
\caption[] {\small Physical parameters for the RRab stars. The numbers in parentheses indicate the uncertainty on the last decimal place.}
\hspace{0.01cm}
 \begin{tabular}{cccccccccc}
\hline 
Star&[Fe/H]$_{\rm ZW}$  &[Fe/H]$_{\rm UVES}$ &$M_V$ & log~$T_{\rm eff}$  & log~$(L/{\rm L_{\odot}})$ & $D$ (kpc) & $M/{\rm M_{\odot}}$&$R/{\rm R_{\odot}}$&$D_m$\\
\hline
  &  &  & RRab  & &  &  &  &   \\
\hline
\hline
V5 & $-1.63(4)$ & $-1.56(5)$ & 0.538(4) & 3.806(9) & 1.698(2)& 9.13(2) & 0.72(8) & 5.79(1)& 3.0\\
V6$^a$ & $-1.34(7)$& $-1.22(7)$& 0.447(6)& 3.797(7) & 1.727(2)& 9.85(3) & 0.67(6) & 6.24(2)& 4.3\\
V10 & $-1.65(7)$ & $-1.60(9)$& 0.401(6) & 3.792(7) & 1.754(2)& 10.85(3) & 0.70(6) & 6.59(2)&2.2 \\
\hline
Weighted Mean & $-1.63(2)$ & $-1.57(4)$ & 0.488(3) & 3.797(1) & 1.719(1) & 9.67(1) & 0.70(4) & 6.03(1)\\
$\sigma$&  $\pm$0.02  &  $\pm$0.02  &$\pm$0.057&$\pm$0.006&$\pm$0.022 & $\pm$0.70 &$\pm$0.02&$\pm$0.32\\
\hline 
  &  &  & RRc  &  &  &  &  &   \\
\hline
V7$^b$ & $-2.10(--)$ & $-2.25(--)$ & 0.456(33) & 3.813(6) & 1.718(13)& 10.04(15) & 0.76(6) & 5.75(9)&-- \\
V11$^b$ & $-1.01(--)$ & $-0.91(--)$ & 0.590(19) & 3.886(2) & 1.644(8)& 9.69(9) & 0.41(1) & 3.84(3)&-- \\
V12 & $-1.63(37)$ & $-1.57(44)$ & 0.496(18) & 3.886(1) & 1.701(7)& 10.17(9) & 0.41(1) & 4.02(3)&-- \\
V16 & $-1.41(--)$ & $-1.30(--)$ & 0.728(29) & 3.900(4) & 1.609(12)& 8.59(12) & 0.30(2) & 3.34(5)&-- \\
\hline
Weighted Mean & $-1.62(36)$ & $-1.55(41)$ & 0.56(1) & 3.886(1) & 1.676(4) & 9.60(5) & 0.38(1) & 3.79(2)\\
$\sigma$&  $\pm$0.39  &  $\pm$0.49  &$\pm$0.10&$\pm$0.034&$\pm$0.042 & $\pm$0.62 &$\pm$0.18&$\pm$0.90 \\
\hline 
\end{tabular}

\center{$a$: This variable is not included in the calculation of [Fe/H] since its $D_m$ parameter is larger than 3.0.\\
$b$: The light curves of these RRc variables seem incompatible with those of the calibrators used to set the [Fe/H] calibration as they\\ produce non-sense medium values and uncertainties. They were con considered for the overall evaluation of the cluster metallicity.} 

\label{fisicos}
\end{table*}

\subsection{The distance to NGC 6139}

The distance to NGC 6139 has been estimated using several approaches from at least the last 25 years, mostly using the CMD main sequence and the RGB fitting. Being the CMD of this cluster scattered, the distances range from 8.7 to 12 kpc. The average of these estimations is $10.35 \substack{+ 0.463  \\ -0.441}$ kpc \citep{Baumgardt2021}.

We have estimated the cluster distance from the light curve Fourier decomposition and absolute magnitude calibrations (see \S \ref{sec:RRLstars}) and the independent values from RRab and RRc are 9.67 $\pm$0.70 and 9.60 $\pm$0.62 kpc respectively.

Yet an independent estimate can be obtained from the Period-Luminosity relation in the $I$-band from \citet{Catelan2004}; $M_I=0.471-1.132 \log P + 0.205 \log Z$, where $P$ is the fundamental period, $\log Z = {\rm [M/H]} - 1.765$, ${\rm [M/H] = [Fe/H]} - \log(0.638f + 0.205)$ and $\log f = {\rm [\alpha/Fe]}$ \citep{Salaris1993}. For three RRab stars, V5, V6 and V10, and three RRc stars, V11, V12 and V16 (whose periods  were fundamentalized before the application of the calibration by using the ratio $P_1/P_0=0.749$), we found an average distace of $9.67\pm 0.30$ kpc, in a good agreement with the Fourier results. Our results tend to favour a distance a bit shorter than the average of  \citet{Baumgardt2021}, although in agreement within the respective uncertainties.

\section{Summary and conclusion}
\label{summary}

A differential image analysis of a time series collection of 360 and 444 CCD images in $V$ and $I$ respectively, enabled the calculation of 4607 point sources, and a membership analysis on the astrometric data available in $Gaia$-DR3 identified 1106 likely cluster members. The variable stars in the field of NGC 6139 can be grouped as those listed in the Catalogue of Variable Stars in Globular clusters \citep{cle01}, a group of Variables detected by $Gaia$ and a group of newly found in the present work. Their membership status, new variable names assigned to cluster members and their variable type classifications is summarized in Table \ref{variables}.

The Fourier decomposition of the light curves of member RR Lyrae stars enabled the calculation of the mean metallicity and distance of the cluster as [Fe/H]=$-1.63 \pm 0.20$ dex, and $9.60 \pm 0.68$ kpc respectively. A comparison with previous results extracted from the literature is given in Table \ref{comparison}.

\begin{table}
\scriptsize
\caption{Comparison of results for NGC 6139 with previous works.}
\centering
\begin{tabular}{lccc}
\hline
Author & E(B-V) & Distance & [Fe/H] \\
       & mag  & kpc & dex \\
\hline
\citet{Zinn1980} & 0.78 &  & -1.67 \\
\citet{Bica1983} & 0.68 &  & -1.28 \\
\citet{Bica1986} & 0.70 &  & -1.5 \\
\citet{Hazen1991} & 0.74 &  &  \\
\citet{Samus1996} & 0.82-0.87 &  & $\leq$ 2.0 \\
\citet{Zinn1998} & 0.76 & 10 & -1.71 \\
\citet{Ortolani1999} & 0.77 $\pm$ 0.06 & 9.4 $\pm$ 1 &  \\
\citet{Harris1996} (Ed. 2010) & 0.75 & 10.1 & -1.65 \\
\citet{Schlafly2011} & 0.786 $\pm$ 0.017  &  &  \\
\citet{Baumgardt2021} &  & 10.04 $\pm$ 0.45 & \\
this work & 0.786 $^*$ & 9.63 $\pm$ 0.68 & -1.63 $\pm$ 0.20 \\
\hline
\end{tabular}
\label{tab:literature}

\center{$*$: Adopted from \citet{Schlafly2011}} 
\label{comparison}
\end{table}

\begin{figure*}
\includegraphics[width=17.0cm]{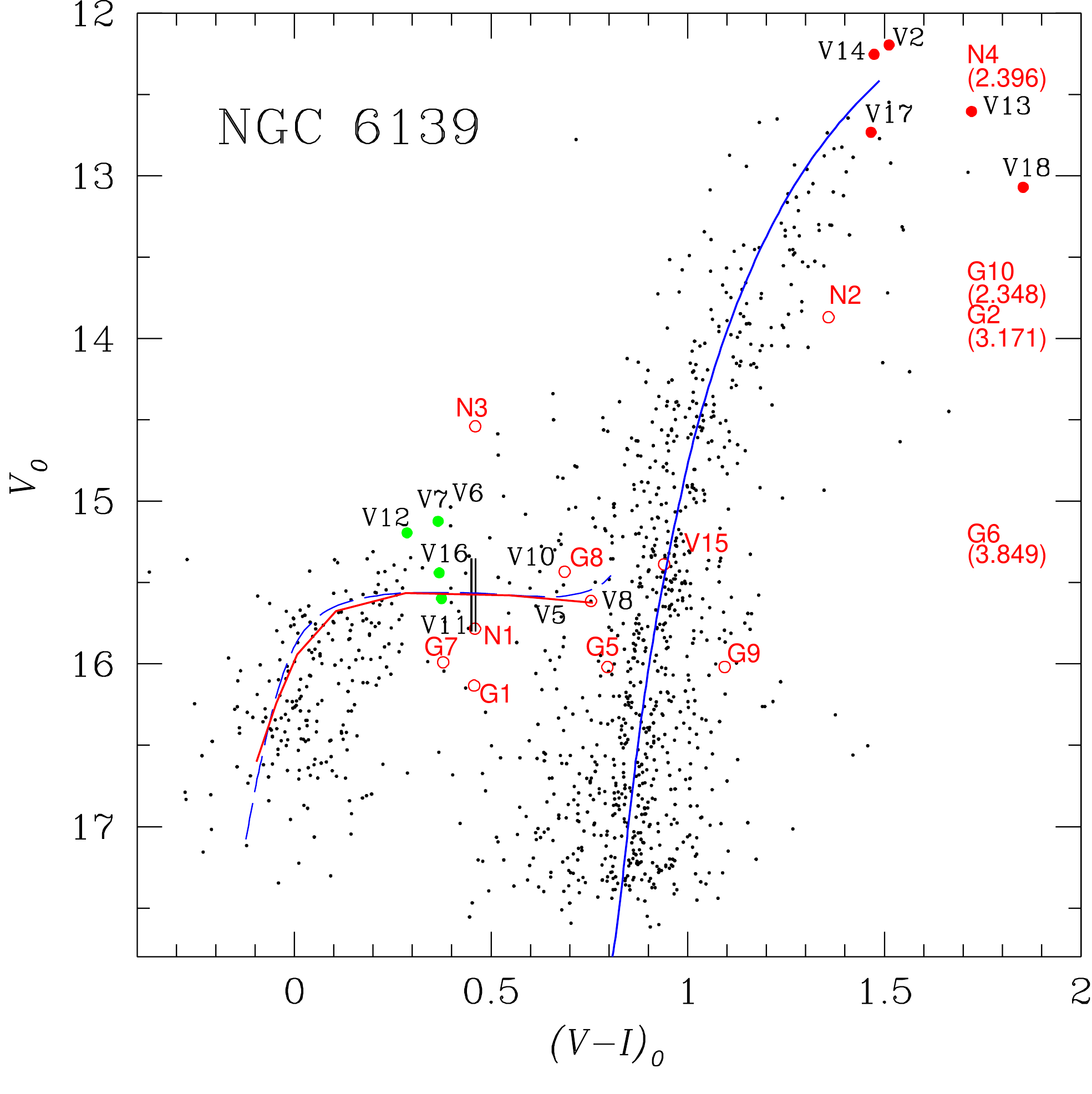}
\caption{CMD of NGC 6139. The diagram has been deferentially dereddened using the map obtained in \S \ref{sec:redd}. Small black dots are likely cluster member stars. Bigger dots represent all variable stars, either previously known or discovered in this work. The labels follow the code: solid blue and green are used for member RRab and RRc stars, solid red for member SR stars, empty circles are for likely field variables. Here 'G' is employed for confirmed variables reported in $Gaia$-DR3 while V15 and 'N' are newly discovered likely field variables. Stars G2, G6, G10 and N4 are very red and likely field stars. Their colour is given in parenthesis.The loci are theoretical predictions from the models of \citet{Vandenberg2014} (blue isochrone and ZAHB) and from the Eggleton code \citep{Pols1997, Pols1998, KPS1997}, calculated by \citet{Yepez2022} (red ZAHB). These theoretical loci have been placed for a mean reddening $E(B-V)=0.786$ and a distance of 9.63 kpc. The vertical black lines at the ZAHB mark the empirical red edge of the first overtone instability strip \citep{Arellano2015,Arellano2016}. The RRab star V6 might indeed be in the fundamental-first overton bimodal region.}
\label{CMD}
\end{figure*}

Being NGC 6139 a cluster near the Galactic bulge, differential reddening is rather expected. Mild differential reddening was found and mapped, which produces a slight improvement in the CMD diagram appearance (see Fig \ref{CMD}). A theoretical isochrone for [Fe/H]=-1.65 dex, $Y=0.25$ and an age of 12 Gyrs and a zero age horizontal branch (ZAHB) (blue lines in Fig \ref{CMD}) \citep{Vandenberg2014} are included in the diagram. For comparison we also included a ZAHB calculated by \citep{Yepez2022} for a core mass of 0.5 $M_{\odot}$ from Eggleton code \citep{Pols1997, Pols1998, KPS1997} (red line in Fig \ref{CMD}). It was found that all these loci represent the observations best if they are shifted to a distance of 9.6 kpc which in a way represents an independent distance estimate.

\section*{Acknowledgments}
We are grateful to Drs. Jesús H. Calderón, Javier Ahumada and Nidia Morrell for their help in acquiring some observations at EABE and Swope.  A. L. acknowledges support from the ANID Doctorado Nacional 2021 scholarship 21211520, and the ESO studentship. The financial support from DGAPA-UNAM (Mexico) via grant IG10062 is acknowledged. We have made an extensive use of the SIMBAD and ADS services, for which we are thankful. Very useful comments and suggestions from an anonymous referee are gratefully recognized.

This work has made use of data from the European Space Agency (ESA) mission {\it Gaia} (https://www.cosmos.esa.int/gaia), processed by the {\it Gaia} Data Processing and Analysis Consortium (DPAC, https://www.cosmos.esa.int/web/gaia/dpac/consortium). Funding for the DPAC has been provided by national institutions, in particular the institutions
participating in the {\it Gaia} Multilateral Agreement.

\appendix
\section{Comments on peculiar variable stars}
\label{sec:comments}

In this section, we include a brief comment on variables with some peculiarities or with unusual positions on the CMD.

V2. Our purely astrometric membership analysis indicates that this is a field star. However, the star is at the tip of RGB and, based on its period, we can classify V2 as a cluster member SR star.

V8. The membership analysis suggests that this contact binary is a field star.

V14. This variable is a new discovery in this paper. It was found  by blinking the residual images. Although its light curve shows a small variation, more data would be required to confirm its variability. The position of V14 in the CDM indicates that it is likely a cluster member SR star.

V15. The period and light curve of this newly found variable suggest it is a RRc star. However, in the CMD, it is located to the red of the HB, on RGB, which could indicate that the star is not member of the cluster subject to larger reddening. The astrometric membership analysis however, suggests that the star belongs to the cluster. Our differential reddening correction did not solve the peculiar position.

V18. The membership analysis indicates that the star is a cluster member. Its light curve and period are typical of a SR star. Its location near the red tip of RGB, confirms this classification. However, like V15, its very red position may be explained by a large reddening.

N1. This star is found to be  a likely field star. However, the star is located on the HB, We found two possible periods: 0.537526 and 0.268666 d, which produced light curves typical of a contact binary and RRc, respectively (see Fig. \ref{newvar}). Given the quality of our light curve, the two scenarios are equally possible.

N2 and N3. Their light curves and periods indicate these are an RRab and RRc stars, respectively. Nevertheless, the membership analysis suggests they are field stars. Their odd position on the CMD confirm this. Given that they do not belong to the cluster, we refrained from assigning a prefix 'V' variable name for them.

N4. Its $V$ and $I$ light curves show clear and similar variations although its period of about 25 days may need refinement with further long time span observations. Its position on the CMD far to the red of the RGB and the membership analysis suggest that it is a field star.

\section*{DATA AVAILABILITY}
The data underlying this article shall be available in an electronic
form in the Centre de Donnés astronomiques de Strasbourg data
base (CDS), and can also be shared on request to the corresponding
author.

\bibliographystyle{mnras}
\bibliography{NGC6139}

\bsp	
\label{lastpage}
\end{document}